\documentstyle[aps,twocolumn,prb]{revtex}
\begin{document}

\draft

\title{ Equilibrium spin current through the tunnelling junctions}

\author{J. Wang and K.S. Chan}
\address{$^{1}$Department of Physics and Materials Science, City University of Hong Kong, Tat Chee
Avenue, Kowloon, Hong Kong, P.R. China}

\date{\today}
\maketitle

\begin{abstract}
We study equilibrium pure spin current through tunnelling
junctions at zero bias. The two leads of the junctions connected
via a thin insulator barrier, can be either a ferromagnetic metal
(FM) or a nonmagnetic high-mobility two-dimensional electron gas
(2DEG) with Rashba spin orbital interaction (RSOI) or Dresselhaus
spin orbital interaction (DSOI). As a lead of a tunnelling
junction, the isotropic RSOI or DSOI in 2DEG can give rise to an
average effective planar magnetic field orthogonal or parallel to
the current direction.  It is found by the linear response theory
that equilibrium spin current $\vec{J}$ can flow in the following
three junctions, 2DEG/2DEG, 2DEG/FM, and FM/FM junctions, as a
result of the exchange coupling between the magnetic moments,
$\vec{h}_{l}$ and $\vec{h}_{r}$, in the two electrodes of the
junction, i.e., $\vec{J}\sim\vec{h}_{l}\times\vec{h}_{r}$. An
important distinction between the FM and 2DEG with RSOI (DSOI)
lead is that in a strict one-dimensional case RSOI (DSOI) cannot
lead to equilibrium spin current in the junction since the two
spin bands are not spin-polarized as in a FM lead where Zeeman
spin splitting occurs.

\end{abstract}

\section{introduction}
Spin related transport in magnetic microstructures has drawn
considerable interests in research community for the purpose of
spintronics.\cite{1,2,3} In order to manipulate the spin degree of
freedom of an electron, one needs first to build a source of
spin-polarized electrons and efficiently injects spins into a
non-polarized medium. The ferromagnetic metal (FM) is an ideal
spin source for spin injection into semiconductor devices;
however, the injection efficiency experimentally measured is
extremely low due to the much larger conductivity of FM than that
of semiconductor.\cite{4,5,6} This issue has not yet been resolved
although it is fundamentally important. As an alternative solution
to the spin injection problem, the pure spin current was proposed
and has attracted much attention in recent years. The pure spin
current is composed of equal spin down and spin up electron
currents flowing along opposite directions without any charge
current.\cite{7} Several different methods have been proposed to
create pure spin current. Based on the spin pumping
principle,\cite{8,9} an alternating or inhomogeneous magnetic
field as a spin-pumping force could result in a pure spin current.
Experimentally, Stevens \emph{et al.}\cite{10} and Hubner \emph{et
al.}\cite{11} have independently realized the pure spin current by
using the quantum interference of two-color laser fields with
cross-linear polarization in ZnSe and GaAs semiconductors. The
transport of spin current by magnons have been theoretically
studied by several groups,\cite{12,13,14} e.g., Meier \emph{et
al.}\cite{12} demonstrated that by using a finite length spin
chain between magnetic reservoirs, pure spin current can be
generated without the transport of electrical charge.

The equilibrium spin current (ESC) in a magnetic or nonmagnetic
system is also very attractive. The newly unearthed spin Hall
current, discovered by Murakami \emph{et al.}\cite{15} in
$p$-doped semiconductors and Sinova \emph{et al.}\cite{16} in
two-dimensional electron systems (2DEG),  is actually a
dissipationless spin current, which is a transverse response to a
longitudinal external electric field $E_{x}$. The spin Hall
current in a 2DEG is generated by the Rashba spin orbital
interaction (RSOI). In a single RSOI system the spin state of an
electron is dependent on its momentum direction, so that at
equilibrium the spin current could flow in the system without
bias.\cite{17} As long as a system has noncollinear magnetic
order, a ESC could exist in the system.\cite{18} For instance,
K{\"{o}}{nig} \emph{et al.}\cite{19} found a dissipationless spin
current in a thin film ferromagnet, in which a spiral magnetic
order exists and the spin phase coherence can affect the
electronic transport properties.

For the FM/FM tunnelling junction, a spontaneous ESC has been
verified\cite{20,21,22,23} flowing across the interface when the
two magnetic moments $\vec{h}_{l}$ and $\vec{h}_{r}$ in the two
leads of the junction are not collinear, and their exchange
interaction determines the ESC $\vec{J}_s$,
$\vec{J}_{s}\sim\vec{h}_{l}\times\vec{h}_{r}$. This indicates
non-collinear magnetizations in the two leads can lead to spin
flip and consequently a ESC through the junction. Nogueira
\emph{et al.}\cite{22} and Lee \emph{et al.}\cite{23} referred to
the ESC through the FM/FM junction as the spin supercurrent as an
analogy to the Josephson effect in superconductor junctions. The
spin current in a FM/FM junction was also shown to result in
magnetic moment reversal in one of the FM leads of the
junction.\cite{24}

In 2DEG with RSOI or Dresselhaus spin orbital interaction (DSOI),
there is a pseudomagnetic field which could lead to ESC in the
non-magnetic system. This pseudomagnetic field is different from a
real magnetic field as it is dependent on the direction of
electron momentum and keeps the system's time reversal symmetry. A
natural question thus arises whether a spin supercurrent (or ESC)
could flow through a \emph{nonmagnetic} tunnelling junction in
which the electrodes are 2DEGs with either RSOI or DSOI. A recent
paper\cite{25} by B{\o}rkje and Sudb{\o} has been dedicated to
this issue and authors obtained a ESC in a 2DEG junction with RSOI
by rotating one of the 2DEG electrodes along the current direction
to change the direction of the pseudomagnetic field, so that the
product of two pseudomagnetic fields in two leads is nonzero,
meanwhile, the absence of ESC through the 2DEG/FM tunnelling
junction was also predicted. In this paper, we will study
systematically the ESC in three different junctions: 2DEG/2DEG,
2DEG/FM, and FM/FM junctions, using the linear response theory. It
is found that ESC is present in all three junctions resulting from
the exchange coupling between the two magnetic moments in the two
electrodes of the junctions. Non-magnetic 2DEG with RSOI or DSOI
as an electrode in a tunnelling junction can lead to a ESC,
because only half of the electrons (those with $k_{x}>0$, if the
2DEG is on the left of the junction) in an electrode contribute to
the tunnelling current so that the isotropic RSOI or DSOI can give
rise to an average planar effective magnetic field orthogonal or
parallel to the current direction. We also found that in a strict
one-dimensional (1D) case RSOI or DSOI cannot result in a ESC in
the junction because the 1D density of states is not
spin-polarized, which is very different from that of a 1D FM lead.

This paper is organized in the following way. In the second part,
we give a general formula of the ESC in  a tunnelling junction
derived using the linear response theory. In the third part, we
analyze in detail the ESC in three junctions: 2DEG/2DEG, 2DEG/FM,
and FM/FM junctions. A conclusion was drawn in the last section.

\section{formula}
We start our discussion with the derivation of a general formula
of ESC through a junction, consisting of two leads of either 2DEG
with RSOI (DSOI) or FM metal and an insulator barrier between
them. The Hamiltonian of such a tunnelling system is given by
\begin{mathletters}
\begin{eqnarray}
{\cal H}={\cal H}_{L}(C^{\dagger}_{k\sigma};C_{k\sigma})+{\cal
H}_{R}(C^{\dagger}_{q\sigma};C_{q\sigma})+{\cal H}_{T}
\end{eqnarray}
\begin{eqnarray}
{\cal
H}_{T}=\sum_{kq\sigma}(t_{kq}C_{k\sigma}^{\dagger}C_{q\sigma}+\emph{h.c.}),
\end{eqnarray}
\end{mathletters}
where ${\cal H}_{L(R)}$ is the Hamiltonian of a non-interacting
free-electron gas in the left (right) lead, whose specific form
will be present in next section; ${\cal H}_{T}$ is the tunnelling
part connecting two leads. $C_{k\sigma}^{\dagger}(C_{k\sigma})$
and $C_{q\sigma}^{\dagger}(C_{q\sigma})$ are the fermion creation
(annihilation) operators of the left and right leads,
respectively. The hopping matrix $t_{kq}$ is independent of spin,
so that there is no spin-flip when electrons tunnel through the
junction. In the following discussion the quantum number $k$ and
$q$ can also denote implicitly the left and right lead.

We focus on the tunnelling ESC and the bulk ESC in the RSOI (DSOI)
lead is not considered here because it does not contribute to the
tunnelling current. The total spin in the left lead is
${\mathbf{S}}_{L}=(\hbar/2)\sum_{k\alpha\beta}C_{k\alpha}^{\dagger}{
{{\sigma}}_{\alpha\beta}}C_{k\beta}$ where $\alpha$ and $\beta$
are spin indices, and ${{\sigma}}$ is the Pauli spin operator. The
time evolution of ${\mathbf{S}}_{L}$ in the Heisenberg picture is
given by
$\dot{\mathbf{S}}_{L}=(1/i\hbar)[{\mathbf{S}}_{L},H_{T}]$, and
thus the operator of spin current
$\hat{{\mathbf{J}}}_{s}=\partial{{\mathbf{S}}_{L}}/\partial{t}$
reads
\begin{equation}
{\hat{\mathbf{J}}}_{s}=\frac{-i}{2}\sum_{kq\alpha\beta}{{\sigma}}_{\alpha\beta}
(t_{kq}C_{k\alpha}^{\dagger}C_{q\beta}-t_{kq}^{*}C_{q\alpha}^{\dagger}C_{k\beta}).
\end{equation}
Here the coupling ${\cal H}_{T}$ is assumed to be very weak and we
only need to calculate the spin current to the lowest order of the
coupling. Since the spin current operator $\hat{\mathbf{J}}_{s}$
itself is already linear in $t_{kq}$, and thus according to the
Kubo formula the spin current is to first order in ${\cal{H}}_{T}$
given by
\begin{mathletters}
\begin{eqnarray}
{{\mathbf{J}}}_{s}(t)=\int_{-\infty}^{t}dt^{\prime}e^{-0^{+}(t-t^{\prime})}\Pi_{R}(t,t^{\prime}),
\end{eqnarray}
\begin{eqnarray}
\Pi_{R}(t,t^{\prime})=-i\theta(t-t^{\prime})\langle[\hat{{\mathbf{J}}}_{s}(t),{\cal
H}_{T}(t^{\prime}) ]\rangle,
\end{eqnarray}
\end{mathletters}
where $\langle\cdots\rangle$ is the thermal statistical average on
two independent leads ${\cal H}_{L}+{\cal H}_{R}$ (the unperturbed
part of $\cal{H}$), which are assumed to be in local equilibrium.
$\Pi_{R}(t,t^{\prime})$ is the retarded Green's function. After
some straightforward algebra and keeping the average only in lead
${\cal H}_{L}$ or ${\cal H}_{R}$, we could express Eq.~3 in the
steady state as
\begin{eqnarray}
{\mathbf{J}}_{s}=\frac{-1}{2}\int\frac{d\omega_{1}}{2\pi}\frac{d\omega_{2}}{2\pi}
\sum_{kq}|t_{kq}|^{2}Tr\{ \nonumber \\
 \frac{G_{k}^{<}(\omega_{1}){{\sigma}}
G_{q}^{>}(\omega_{2})-G_{k}^{>}(\omega_{1}){{\sigma}}
G_{q}^{<}(\omega_{2})}{i(\omega_{2}-\omega_{1}-i0^{+})}
+\nonumber \\
\frac{G_{q}^{>}(\omega_{2}){{\sigma}}
G_{k}^{<}(\omega_{1})-G_{q}^{<}(\omega_{2}){{\sigma}}
G_{k}^{>}(\omega_{1})}{i(\omega_{1}-\omega_{2}-i0^{+})}
 \},
\end{eqnarray}
where the trace $Tr$ is over the spin space, $G_{k(q)}^{<(>)}$ are
the Keldysh Green's functions in the left (right) lead and can be
readily solved for leads of free-electron model. Their
definitions are $G_{k(q),\alpha\beta}^{<}(t,t^{\prime})=i\langle
C_{k(q)\beta}^{\dagger}(t^{\prime})C_{k(q)\alpha}(t)\rangle$ and
$G_{k(q),\alpha\beta}^{>}(t,t^{\prime})=-i\langle
C_{k(q)\alpha}(t) C_{k(q)\beta}^{\dagger}(t^{\prime})\rangle$,
respectively. The Fourier transform of the Green's function is given
by $G_{k(q)}^{<(>)}(t-t^{\prime})=
\int\frac{d\omega}{2\pi}G_{k(q)}^{<(>)}(\omega)e^{-i\omega(t-t^{\prime})}$.

It is noted that Eq.~4 above is a general formula of spin current
through a tunnelling junction and the bias could be reflected in
the local equilibrium Green's functions $G_{k(q)}^{<(>)}$ of the
two leads. For zero bias on the junction considered in this
article, i.e., zero charge current flow through the junction, we
can further simplify Eq.~4 by putting
$1/(x+i0^{+})={\sf{P}}(1/x)-i\pi\delta(x)$ with $\sf{P}$ denoting
the principle integral,
\begin{mathletters}
\begin{eqnarray}
{\mathbf{J}}_{s}=\frac{1}{2}\int\frac{d\omega_{1}}{2\pi}\frac{d\omega_{2}}{2\pi}
\sum_{kq}|t_{kq}|^{2}{\sf{P}}\frac{f(\omega_{2})-f(\omega_{1})}{i(\omega_{2}-\omega_{1})}
Tr\{ \nonumber \\
A_{k}(\omega_{1}){{\sigma}}
A_{q}(\omega_{2})-A_{q}(\omega_{2}){{\sigma}}A_{k}(\omega_{1}) \},
\end{eqnarray}
\begin{eqnarray}
A_{k(q)}(\omega)=i(G^{r}_{k(q)}(\omega)-G^{a}_{k(q)}(\omega)).
\end{eqnarray}
\end{mathletters}
Here $A_{k(q)}(\omega)$ is the spectral function of the left
(right) lead with $G^{r(a)}(\omega)$ denoting the retard
(advanced) Green's function and $f(\omega)$ denoting the Fermi
distribution function. The ESC in Eq.~5 is mainly determined by
the difference between two Hermitian-conjugate matrices in spin
space, so that the off-diagonal term in the spectral function
$A_{k(q)}$ is utmost important to the formation of a ESC in a
tunnelling junction. For FM or 2DEG with RSOI (DSOI), this
off-diagonal term in spin space exists when two magnetizations
(pseudomagnetic fields in 2DEG) are not collinear in the two
leads, as a result, a nonzero ESC can flow through the junction
according to Eq.~5. In the following section we will discuss the
ESC in three specific junctions: 2DEG/2DEG, 2DEG/FM, and FM/FM
junctions.

\section{results and discussion}
\subsection{2DEG/2DEG tunnelling junction}
In this section we study the ESC through a 2DEG/2DEG junction with
spin orbital interaction. First we review briefly the
characteristics of a 2DEG with RSOI, which stems from the
structure inversion asymmetry of the confining potential of a
quantum well.\cite{26} The magnitude of the RSOI strength could be
modulated by a vertical electric field $E\vec{z}$ ($\vec{z}$ is
the unit vector in the $z$ direction in Fig.~1) applied on the
2DEG. The Hamiltonian of the 2DEG with RSOI is
\begin{mathletters}
\begin{eqnarray}
{\cal{H}}=\frac{\mathbf{p}^{2}}{2m_s}+{\cal{H}}_{rsoi},
\end{eqnarray}
\begin{eqnarray}
{\cal{H}}_{rsoi}=\frac{\lambda}{\hbar}(\sigma_y p_x-\sigma_x p_y),
\end{eqnarray}
\end{mathletters}
where $m_s$ is the effective mass of electrons in the 2DEG, $p_x$
and $p_y$ are the two components of the momentum operator
$\textbf{p}$, and $\lambda$ is the coupling constant and was
estimated to be $\sim 3.9\times 10^{-12}$ eV m in an InGaAs/InAlAs
heterostructure.\cite{27} $\sigma_x$ and $\sigma_y$ are the Pauli
matrices, and the $z$-axis is taken as the spin quantum axis.

The electron eigenfunctions of Eq.~(6) are
$\psi_{\pm}({\mathbf{r}})=(1/\sqrt{2})e^{i{\mathbf{k}\cdot
r}}\left( \begin{array}{c}
  \mp ie^{-i\theta_{r}} \\
 1 \\
\end{array}\right)$
with $\tan\theta_{r}=k_{y}/k_{x}$, and the corresponding
eigenvalues are $\varepsilon_{\pm}=\hbar^{2}k^2/2m_{s}\pm\lambda
k$. This wavefunction of electron reminds us that the spin orbital
interaction can give rise to a magnetic field parallel to the
plane, which is dependent on the direction of the electron
momentum. From Eq.~6b, the effective (pseudo) magnetic field is
$\vec{B}\sim \vec{z}\times \mathbf{p}$. Different from a real
magnetic field, the RSOI keeps the system's isotropy and
time-reversal symmetry $\mathbf{B(k)}=\mathbf{-B(-k)}$, and the
2DEG does not exhibit a real spin splitting like in the Zeeman
effect. This will result in a distinction between the ESC in a
2DEG junction and the ESC in a FM junction.

Although the pseudomagnetic field $B$ in 2DEG is isotropic, only
half of the electrons (electrons with $k_{x}>0$ on the left and
$x$ is the current direction in Fig.~1) in the 2DEG contribute to
the tunnelling current, as a result, the average pseudomagnetic
field of the tunnelling electrons is nonzero and along the $y$
direction ($B\vec{y}$). If we could change the direction of the
pseudomagentic field in one of the 2DEG, a ESC would flow through
a 2DEG/2DEG junction. A celebrated example is the RSOI/DSOI
junction, where one 2DEG lead possesses RSOI while the other has
DSOI. By exchanging the spin axis
$\sigma_{x}\longleftrightarrow\sigma_{y}$, RSOI can be transformed
into DSOI as
\begin{equation}
{\cal{H}}_{dsoi}=\frac{\lambda}{\hbar}(\sigma_x p_x-\sigma_y p_y).
\end{equation}
This coupling is due to the lack of the bulk inversion symmetry in
the material.\cite{28} Thus in the RSOI/DSOI tunnelling junction,
the average pseudomagnetic field in the DSOI lead is along the $x$
direction, which is different from that in the RSOI lead, so that
a nonzero ESC $\sim B_{rsoi}\vec{y}\times B_{dsoi}\vec{x}$ exists
as shown below in Eq.~11. In Ref.~25, B{\o}rkje and Sudb{\o}
attempted to rotate one 2DEG lead along the current direction
($x$-direction in Fig.~1) so as to rotate the orientation of the
pseudomagnetic field. However this method may not create a ESC
through the RSOI/RSOI junction as discussed below. The rotation of
the average psuedomagnetic field could be also obtained by
rotating physically a quasi-one-dimensional 2DEG along the
$z$-axis by an azimuthal angle $\alpha$, which is schematically
shown in Fig.~1. Since the pseudomagnetic field $B\sim
\vec{z}\times \mathbf{p}$ is dependent on the direction of the
electron momentum, an oblique incident electron will feel
pseudomagentic fields with different orientations when it tunnels
through a RSOI/RSOI junction where the energy band edges of the
two 2DEG leads are different. This change in psuedomagnetic field
direction could lead to a ESC in the junction. The energy band
shift in one 2DEG lead may be obtained by applied a gate voltage
or mechanical strain to the lead.

To understand the relation of a rotated pseudomagnetic field and
ESC, we consider a rotation of the spin $x-y$ axes by an angle
$\alpha$ with respect to the spatial $x-y$ axes. The rotated angle
$\alpha$ of the pseudomagnetic field could be
incorporated into the Pauli matrices $\sigma_x(\alpha)=\left(%
\begin{array}{cc}
  0 & e^{i\alpha} \\
  e^{-i\alpha} & 0 \\
\end{array}%
\right)$ and $\sigma_y(\alpha)=\left(%
\begin{array}{cc}
  0             & -ie^{i\alpha} \\
  ie^{-i\alpha} & 0              \\
\end{array}%
\right)$. Here $\alpha$ is macroscopic in that it applies to all
the electrons and may denote the rotation angle of the average
pseudomagnetic field.

In second quantization formalism, Eq.~6 can be rewritten as
\begin{equation}
{\cal{H}}=\sum_{k}\left(%
\begin{array}{cc}
  C_{k\uparrow}^{\dagger} & C_{k\downarrow}^{\dagger} \\
\end{array}%
\right)
\left(%
\begin{array}{cc}
  \varepsilon_{k} & -i\lambda ke^{i\theta} \\
  i\lambda ke^{-i\theta} & \varepsilon_{k} \\
\end{array}%
\right)
\left(%
\begin{array}{c}
  C_{k\uparrow} \\
  C_{k\downarrow} \\
\end{array}%
\right),
\end{equation}
where $\varepsilon_{k}=\hbar^2k^2/2m_{s}-\mu$ with $\mu$ the
chemical potential, and $e^{i\theta}=e^{i(\alpha-\theta_{r})}$.
With $\theta_{r}$ replaced by $\theta_{d}$
($\tan\theta_{d}=k_{x}/k_{y}$) as well as $\lambda$ by $-\lambda$,
this Hamiltonian can represent a free 2DEG with DSOI in Eq.~7. The
retard (advanced) Green's function of this free electron model
with RSOI is easily given by
\begin{equation}
G^{r(a)}(\omega)=\left(%
\begin{array}{cc}
  \omega\pm i0^{+}-\varepsilon_{k}& i\lambda ke^{i\theta} \\
  -i\lambda ke^{-i\theta} & \omega\pm i0^{+}-\varepsilon_{k} \\
\end{array}%
\right)^{-1}.
\end{equation}
Before we present the final expression of ESC of 2DEG/2DEG junction by
substituting $G^{r(a)}$ above into Eq.~5, we first focus on the
spectral function $A_{k}(\omega)$ of 2DEG with RSOI
\begin{equation}
A_{k}=\left(%
\begin{array}{cc}
  \pi [\delta_{+}+\delta_{-}] & -i\pi [\delta_{-}-\delta_{+}]e^{i\theta}   \\
  i\pi [\delta_{-}-\delta_{+}]e^{-i\theta} & \pi [\delta_{+}+\delta_{-}] \\
\end{array}%
\right),
\end{equation}
where $\delta_{+}=\delta(\omega-\varepsilon_k+\lambda k)$ and
$\delta_{-}=\delta(\omega-\varepsilon_k-\lambda k)$ with $\delta$
denoting the delta function. In the formula of ESC (Eq.~5), the
summation term $\sum_{k}A_{k}$ exists in the case of weak
connection, i.e., $t_{kq}=t$ independent of energy. In a
\emph{strict one-dimensional} case, the off-diagonal terms of this
summation vanish $\sum_{k}A_{k}(12)=\sum_{k}A_{k}(21)=0$; thus,
there is no ESC through the junction as discussed earlier (Eq.~5).
This result stems from the fact that the 1D density of states of
the two spin bands of an electron gas with RSOI or DSOI is not
spin-polarized, which is very different from the FM case. In the
1D case the density of states is inversely proportional to the
electron velocity, and the velocities of electrons  in the two
spin bands with the same energy  under RSOI or DSOI are identical
so that the summation of the off-diagonal terms in Eq.~10 over $k$
vanishes. In other words, the perpendicular incidence of electrons
from one RSOI (DSOI) lead into another RSOI (DSOI) lead will
conserve their spins whereas this is not the case for the
obliquely incident electrons.\cite{29} Therefore, the rotation of
a 2DEG along the current direction (the $x$-axis in Fig.~1) in the
2DEG/2DEG tunnelling junction will give rise to two 2DEG leads not
in the same plane, and if the middle insulator barrier is very
thin, the electron tunnelling in such a junction may be a 1D
transport, i.e., only electrons with velocity along the
$x$-direction could tunnel through the junction due to the
momentum mismatch. Thus the ESC may vanish in this scheme.

After straightforward algebra, the ESC for the two-dimensional
RSOI/DSOI junction (two 2DEGs in the same plane) is given by
\begin{mathletters}
\begin{eqnarray}
J_{s}^x=0,
\end{eqnarray}
\begin{eqnarray}
J_{s}^y=0,
\end{eqnarray}
and
\begin{eqnarray}
J_{s}^{z}=\frac{\hbar}{2\pi
e^2}G_{N}E_{r}f(-E_{r})\sin(\alpha_L-\alpha_R-\pi/2).
\end{eqnarray}
\end{mathletters}
Here $\alpha_{L(R)}$ is the rotation angle of the spin axes of the
left (right) lead that changes the direction of the pseudomagnetic
field. $E_{r}=\hbar^{2}k_{r}^{2}/2m_{s}$ is the Rashba energy with
$k_{r}=\lambda m_{s}/\hbar^{2}$, $f(-E_{r})$ is the Fermi
distribution function.
$G_N=\frac{e^{2}}{\hbar}2\pi|t|^{2}\rho_{L}\rho_{R}$ is the
conductance of the normal 2DEG junction without spin-orbital
interaction, $\rho_{L(R)}$ is a constant density of state of 2DEG.
$\pi/2$ phase difference in Eq.~11c comes from the exchange of
spin axes when DSOI is compared with RSOI. Here we have assumed
for simplicity the spin-orbital coupling constants of two leads
are same $\lambda_{L}=\lambda_{R}=\lambda$ and $t_{kq}=t$ is
independent of energy.

From Eq.~11, only ESC polarized along the $z$ direction is nonzero
since the two pseudomagnetic fields in RSOI/DSOI junction lie in
the $xy$ plane, their exchange coupling will lead to spin current
polarized along the $z$ direction. The phase (rotation angle)
$\alpha$ of the pseudomagnetic field is cannonically conjugate to
the spin $s_z$ so that they satisfy the commutation relation
$[s_z,\alpha]=-i$, which means the spin component $s_z$ is not a
conserved quantity as the RSOI (DSOI) can change $s_z$ by rotating
the spin, and ESC could hence flow in these junctions. The physics
is similar to the Josephson current in a superconductor
junction,\cite{22,23} in which the macroscopic phase $\phi$
resulted from the condensation of Cooper pairs is conjugate to the
particle number $N$ and $[\hat{N},\phi]=-i$.\cite{30} The
continuity equation of the spin current in 2DEG with RSOI(DSOI) is
$\partial s/\partial t+\Delta \cdot J_{s}=
 S$, where $J_{s}={\bf Re}\left[ \Psi^{\dagger} \hat{\vec{v}}s\Psi\right]$
is the spin current density with $\hat{\vec{v}}$ being the
velocity operator, and $S={\bf Re}\left[
\Psi^{\dagger}(\vec{B}\times \sigma)\Psi\right]$ is a source term
from the pseudo-magnetic field by rotating the spin component
$s_z$.\cite{31} A similar source term appears also in the
continuity equation of the current of superconductor.\cite{32}

\subsection{2DEG/FM junction}
We turn to discuss ESC in an 2DEG/FM junction. In the last
subsection it was stated that RSOI (DSOI) in 2DEG can be
equivalent to a pseudomagnetic field along the $y$ ($x$) direction
with $\alpha=0$ when electrons tunnel through the junction. When
one of the 2DEG leads is substituted by a FM, a ESC could also
flow through the 2DEG/FM junction from the point of view of the
magnetic field. The Hamiltonian of the FM in the right side of the
junction is given in a simple framework of the Stoner model by
\begin{equation}
{\cal{H}}_{fm}=\sum_{q}\left(%
\begin{array}{cc}
  C_{q\uparrow}^{\dagger} & C_{q\downarrow}^{\dagger} \\
\end{array}%
\right)
\left(%
\begin{array}{cc}
  \varepsilon_{q}+h & 0 \\
  0 & \varepsilon_{q}-h \\
\end{array}%
\right)
\left(%
\begin{array}{c}
  C_{q\uparrow} \\
  C_{q\downarrow} \\
\end{array}%
\right).
\end{equation}
Here the spin quantum axis is set to be along the local magnetic
moment of FM, $\varepsilon_q=\hbar^{2}q^2/2m-\mu$ and $h$ (use
energy as the unit here) is a molecular field in the FM. Then the
local retarded Green's function is
\begin{equation}
g^{r}_{\uparrow(\downarrow)}=\frac{1}{\omega+
i0^{+}-\varepsilon_q\mp h}.
\end{equation}
In the common spin quantum axis set along the $z$-direction for
the 2DEG/FM junction, i.e. the normal of the 2DEG plane, the
Green's functions of the FM lead above should be transformed
according to
\begin{mathletters}
\begin{eqnarray}
G^{r(a)}=U\left(%
\begin{array}{cc}
  g^{r(a)}_{\uparrow} & 0 \\
  0 & g^{r(a)}_{\downarrow} \\
\end{array}%
\right)U^{\dagger},
\end{eqnarray}
\begin{eqnarray}
U=\left(%
\begin{array}{cc}
  \cos\frac{\theta}{2} & -\sin\frac{\theta}{2}e^{-i\phi} \\
  \sin\frac{\theta}{2}e^{i\phi} & \cos\frac{\theta}{2} \\
\end{array}%
\right),
\end{eqnarray}
\end{mathletters}
where $U$ is the unitary transformation matrix, ($\theta$,$\phi$)
denotes the direction of the magnetic moment of the FM lead, which
in terms of the unit vectors of the cartesian coordinates should
be written as ($\sin\theta\cos\phi
\vec{x}$,$\sin\theta\sin\phi\vec{y}$,$\cos\theta\vec{z}$). For the
specific RSOI/FM junction we consider, the rotation angle in the
left lead is set as $\alpha_{L}=0$. Substituting Eq.~14 and Eq.~9
into Eq.~5, we obtain the final result as
\begin{mathletters}
\begin{eqnarray}
J_{s}^{x}=\int
d\omega_{1}d\omega_{2}|t|^{2}{\sf{P}}\frac{f(\omega_{2})-f(\omega_{1})}
{\omega_{2}-\omega_{1}}\times\nonumber \\
\chi_{fm}(\omega_{2})\chi_{2deg}(\omega_{1})\cos\theta
\end{eqnarray}
\begin{eqnarray}
J_{s}^{y}=0
\end{eqnarray}
\begin{eqnarray}
J_{s}^{z}=\int
d\omega_{1}d\omega_{2}|t|^{2}{\sf{P}}\frac{f(\omega_{2})-f(\omega_{1})}
{\omega_{2}-\omega_{1}}\times\nonumber \\
\chi_{fm}(\omega_{2})\chi_{2deg}(\omega_{1})\sin\theta\cos\phi
\end{eqnarray}
\end{mathletters}
with
\begin{mathletters}
\begin{eqnarray}
\chi_{fm}(\omega)=\sum_{q}\{\delta(\omega-\varepsilon_{q}-h)-\delta(\omega-\varepsilon_{q}+h)\}
\end{eqnarray}
\begin{eqnarray}
\chi_{2deg}(\omega)=\sum_{k}
{^\prime}\{\delta(\omega-\varepsilon_k-\lambda
k)-\delta(\omega-\varepsilon_k+\lambda k)\}.
\end{eqnarray}
\end{mathletters}
Here the prime $\prime$ on the summation in Eq.~16b means that the
incident angle $\theta_{r}$ of  electron from 2DEG into FM has
been integrated out from $-\pi/2$ to $\pi/2$, which denotes only
$k_x>0$ electrons contribute to the tunnelling current in our
calculation. $\chi(\omega)$ is the difference of the density of
states of two spin bands in FM and 2DEG. For the 1D case
$\chi_{2deg}=0$ as discussed earlier; whereas for the FM cases
with any dimensions , $\chi_{fm}\neq 0$ if the molecular field $h$
is nonzero. Thus in 1D RSOI(DSOI)/FM junction, ESC does not exist.

Eq.~15 indicates that in a RSOI/FM junction at zero bias a ESC can
also flow through the junction and RSOI in the left lead acts as a
magnetic field with direction along the $y$-direction. According
to the exchange coupling between the two magnetic moments in the
two sides of the junction, $J_{s}^{y}$ is zero while
$J_{s}^{x}\sim\cos\theta$ and $J_{s}^{z}\sim\sin\theta\cos\phi$
are nonzero. If RSOI in the left lead is replaced by DSOI to form
a DSOI/FM junction, the ESC of Eq.~15 is  different,
$J_{s}^{x}=0$, $J_{s}^{y}\sim\cos\theta$, and
$J_{s}^{z}\sim\sin\theta\sin\phi$ because in this case DSOI gives
rise to an average pseudomagnetic field along the $x$-direction.
The results we obtained here indicate again the RSOI (DSOI) in the
tunnelling junction can play the same role as in a FM metal.

\subsection{FM/FM}
For completeness we  present in this subsection the results of the
ESC through a FM/FM junction, which are same as those found in
literatures.\cite{21,22,23}  The spin quantum axis is taken as the
magnetic moment direction of the left FM lead
$\vec{{\mathbf{h}}}_{L}$($0$,$0$,$1$) and the magnetic moment in
the right FM lead is same as that in the last subsection
$\vec{{\mathbf{h}}}_{R}$($\sin\theta\cos\phi $,
$\sin\theta\sin\phi$, $\cos\theta$). Using the Green's function
(Eq.~14) of the FM lead, we can obtain the formula of ESC as
\begin{mathletters}
\begin{eqnarray}
J_{s}^{x}=-\frac{1}{2}\int
d\omega_{1}d\omega_{2}|t|^{2}{\sf{P}}\frac{f(\omega_{2})-f(\omega_{1})}
{\omega_{2}-\omega_{1}}\times\nonumber \\
\chi_{fm}^{R}(\omega_{2})\chi_{fm}^{L}(\omega_{1})\sin\theta\sin\phi,
\end{eqnarray}
\begin{eqnarray}
J_{s}^{y}=\frac{1}{2}\int
d\omega_{1}d\omega_{2}|t|^{2}{\sf{P}}\frac{f(\omega_{2})-f(\omega_{1})}
{\omega_{2}-\omega_{1}}\times\nonumber \\
\chi_{fm}^{R}(\omega_{2})\chi_{fm}^{L}(\omega_{1})\sin\theta\cos\phi,
\end{eqnarray}
\begin{eqnarray}
J_{s}^{z}=0.
\end{eqnarray}
\end{mathletters}
Here $\chi_{fm}^{L(R)}$ is the difference between the density of
states of the two spin bands of the left (right) FM lead as shown
in Eq.~16a. In a simpler form, Eq.~17 can be written as
${\mathbf{J}}_{s}\sim
\vec{{\mathbf{h}}}_{L}\times\vec{{\mathbf{h}}}_{R}$, which means
the ESC comes from the exchange coupling between the two
magnetizations in the left and right FM leads. Only the spin
component along
$\vec{{\mathbf{h}}}_{L}\times\vec{{\mathbf{h}}}_{R}$ is not a
conserved quantitiy since both magnetizations in the two sides of
the junction can flip the spin. The ESC Eq.~11 and Eq.~15 for the
2DEG/2DEG and 2DEG/FM junction can also be expressed as the form
of vector product of (pseudo)magnetic moments.

In earlier works,\cite{24} this spin current in the FM/FM junction
was found to transport spin torque and result in magnetization
reversal in an unfixed FM lead. The spin current can also induce
an electric field, which may be used to experimentally detect the
ESC through magnetic junctions.\cite{12,13} We wish to point out
here the ESC stemming from the exchange coupling of two magnetic
moments in two leads is achieved based on the approximation of the
linear response, beyond which it is unclear whether the above
results are still valid, thus it is worthwhile to study the
behavior of ESC at the limit of the strong coupling.

\section{summary}
We have presented a detailed analysis of the dissipationless ESC
through three tunnelling junctions, 2DEG/2DEG, 2DEG/FM, and FM/FM.
For these three junctions, the ESC comes from the exchange
coupling between two magnetic (or pseudomagnetic) moments in the
two leads in the linear approximation. Although in 2DEG with RSOI
or DSOI, the pseudomagnetic field is isotropic and keeps the time
reversal symmetry, only electrons with $k_{x}>0$ contribute to the
tunnelling current when 2DEG with RSOI or DSOI is used as the left
lead of the junction as in the 2DEG/2DEG and 2DEG/FM junctions, so
that the tunnelling electrons would feel an average nonzero
pseudomagnetic field along the $y$-direction for RSOI or the
$x$-direction for DSOI. The average pseudomagnetic fields have the
same effect as a real magnetic moment and the ESC could flow in
both 2DEG/2DEG and 2DEG/FM junctions. For the RSOI/DSOI junction,
a spontaneous ESC could form since the pseudomagnetic fields
associated with RSOI or DSOI are inherently different. While for
RSOI/RSOI or DSOI/DSOI junctions, one possibility is to rotate one
of the leads to change the orientation of the pseudomagnetic
field, or we could shift the energy band of one of the 2DEG
because the pseudomagnetic field is dependent on the direction of
the electron momentum $\mathbf{P}$, which in turn is determined by
the band edge. This is different from the magnetic moment in a FM
electrode. Another distinction between the \emph{2DEG} with RSOI
(DSOI) and FM lead is that in strict 1D case RSOI (DSOI) cannot
result in a ESC through the junction, while this is not the case
for 1D FM leads.

ACKNOWLEDGEMENT  This work is supported by HongKong Research Grant
Council, Project No: CityU 100303.

\begin{figure}
\caption{Schematic of a tunnelling junction with two quasi-1D
electrodes with RSOI (DSOI) connected by an insulator barrier
(I.B.). The left electrode could be physically rotated along
$z$-direction by angle $\alpha$, which is equivalent to rotating
the direction of the average pseudo-magnetic moment in RSOI
(DSOI). The current flows along $x$-axis. }
\end{figure}

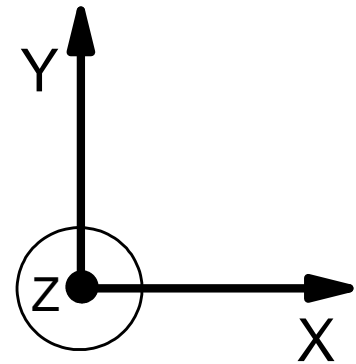
\end{document}